\documentclass{article}

\usepackage[eandd, final]{neurips_2026}

\usepackage[utf8]{inputenc} %
\usepackage[T1]{fontenc}  %
\usepackage{hyperref}    %
\usepackage{url}      %
\usepackage{booktabs}    %
\usepackage{amsfonts}    %
\usepackage{nicefrac}    %
\usepackage{microtype}   %
\usepackage{xcolor}     %
\usepackage{graphicx}    %
\usepackage{multirow}    %
\usepackage{subcaption}   %
\usepackage{todonotes}   %
\usepackage{tabularx}    %
\usepackage{listings}    %
\newcommand{\blue}[1]{\textcolor{blue}{#1}} %
\renewcommand{\blue}[1]{#1} %

\title{ScrapeGraphAI-100k: Dataset for Schema-Constrained LLM Generation}

\author{%
 William Brach\thanks{Equal contribution.} \\
 Slovak University of Technology\\
 Bratislava, Slovakia \\
 \texttt{william.brach@stuba.sk} \\
 \And
 Francesco Zuppichini\footnotemark[1] \\
 ScrapeGraphAI\\
 San Francisco, United States \\
 \texttt{francesco@scrapegraphai.com} \\
 \And
 Marco Vinciguerra \\
 ScrapeGraphAI\\
 San Francisco, United States \\
 \texttt{marco@scrapegraphai.com} \\
 \And
 Lorenzo Padoan \\
 ScrapeGraphAI\\
 San Francisco, United States \\
 \texttt{lorenzo@scrapegraphai.com} \\
}

\begin{document}

\maketitle

\begin{abstract}
Producing output that conforms to a specified JSON schema underlies tool use, structured extraction, and knowledge base construction in modern large language models. Despite this centrality, public datasets for the task remain small, synthetic, or text-only, and rarely pair real page content with the prompts and schemas used in practice. We introduce ScrapeGraphAI-100k, 93{,}695 schema-constrained extraction events collected via opt-out ScrapeGraphAI telemetry in Q2--Q3 2025, deduplicated and balanced by schema from 9M raw events. The corpus spans 25{,}453 unique schemas, across 15 named languages, with English covering 89\% of detected content. Each instance pairs Markdown-converted page content with a prompt, schema, LLM response, and per-example jsonschema-rs structural conformance labels. We characterize structural diversity across the corpus and identify sharp failure thresholds as schema complexity grows. As a case study, a 1.7B student fine-tuned on this data closely tracks the output distribution of its GPT-5-nano teacher, though it still trails a 30B-A3B reference (3.3B active parameters) on schema compliance. We offer this distillation result as preliminary evidence that grounding schema-constrained generation in real practitioner workloads at scale enables training and benchmarking that prior synthetic or text-only corpora could not support. ScrapeGraphAI-100k is publicly available on \href{https://huggingface.co/datasets/scrapegraphai/scrapegraphai-100k}{HuggingFace}\footnote{\url{https://huggingface.co/datasets/scrapegraphai/scrapegraphai-100k}}.

\end{abstract}

\section{Introduction}\label{sec:intro}

Producing output that conforms to a specified schema (JSON, function signatures, or typed records) is a core capability of modern large language models, underlying tool use, structured extraction, document understanding, and knowledge base construction. Progress depends on realistic resources that capture the full extraction four-tuple: unstructured content, natural-language prompts, target schemas, and resulting model outputs. The community currently lacks such a resource.
The task combines (i) web content extraction, which isolates meaningful text from raw HTML, typically emitted as Markdown by downstream tools, and (ii) schema-constrained information extraction, which converts such text into structured records. ScrapeGraphAI-100k operates on the Markdown surface that ScrapeGraphAI users actually feed to LLMs\blue{;} raw HTML and DOM release is deferred to future work (Section~\ref{sec:conclusion}). Existing resources cover only one side of this pipeline, making it hard to study end-to-end extraction quality or failure modes under realistic input distributions.
Just as real-usage dialog datasets such as WildChat~\cite{zhao2024wildchat} and LMSYS-Chat-1M~\cite{zheng2024lmsyschat} have exposed capability and safety phenomena invisible in synthetic data, schema-constrained generation lacks a corresponding real-usage resource. This limits progress on open questions: how output reliability scales with schema complexity, whether small language models can close the gap to frontier models under targeted fine-tuning, and what failure modes (malformed outputs, schema violations, value hallucinations) emerge in production.
ScrapeGraphAI-100k is derived from 9M PostHog telemetry events collected during Q2--Q3 2025\blue{:} real prompts, schemas, web content, and LLM responses from production usage. After cleaning, deduplication, and balancing, we release 93,695 examples with schema-validated outputs and fine-grained complexity metadata. Downstream applications include knowledge distillation into small models, schema induction, and IR-adjacent tasks such as structured indexing.

We make the following contributions:
\begin{itemize}
 \item We release ScrapeGraphAI-100k, a real-world dataset for schema-constrained LLM generation, pairing web-derived content with the prompts, schemas, and model outputs used in production extraction pipelines.
 \item We provide schema-centric diagnostics (validation labels and structural complexity metadata) that expose sharp failure thresholds as task complexity grows.
 \item As a case study of dataset utility, we show that a 1.7B model fine-tuned on our data under knowledge distillation approaches a 30B-A3B MoE (3.3B active parameters) on schema-conformance and key F1, measured against teacher targets.
 \item We release a 100-example human-verified evaluation split, which separates structural metrics from extraction correctness, together with PII, domain-coverage, and frontier-baseline analyses.
\end{itemize}

\section{Related work}\label{sec:related}

Research on web data processing spans three connected areas: large-scale corpora for pre-training, content extraction, and structured information extraction. We review each in turn to motivate the gap addressed by ScrapeGraphAI-100k.

\paragraph{Web-Scale Corpora.}
Large-scale datasets derived from Common Crawl~\cite{commoncrawl}, such as C4~\cite{raffel2020exploring} and FineWeb~\cite{penedo2024the}, provide the token volume that powers modern LLMs. While the Information Retrieval community has adopted hybrid resources such as Istella22~\cite{10.1145/3477495.3531740}, which combine raw text with query-document feature vectors, the web extraction domain lacks a comparable unified benchmark. Pre-training corpora systematically strip HTML tags, visual layouts, and hierarchical metadata, discarding the structural signals required for schema extraction. The resulting data ecosystem is rich in volume but poor in structural ground truth.

\paragraph{Content Extraction.}
To recover some of this lost context, the community has developed benchmarks for Web Content Extraction, the task of isolating the main content of a page from boilerplate (e.g., navigation menus, advertisements). Evaluation remained historically fragmented until Bevendorff et al.~\cite{10.1145/3539618.3591920} consolidated eight human-labeled resources and evaluated 14 extraction systems, finding that heuristic methods often remain competitive with pre-transformer baselines. More recently, Liu et al.~\cite{liu2025drippertokenefficientmainhtml} introduced WebMainBench, a benchmark of over 7,800 human-annotated pages for token-efficient main-content identification. However, WebMainBench targets the HTML-to-text task exclusively, leaving the downstream problem of mapping content to structured schemas unaddressed.

\paragraph{Structured Data Extraction.}
Unlike content extraction, structured extraction maps web segments to specific schema fields (e.g., price, author, date). The SWDE dataset~\cite{hao2011swde} and OpenCeres~\cite{lockard-etal-2019-openceres} have served as standard benchmarks, but both are small-scale and do not reflect modern dynamic DOM structures. CrossNER~\cite{liu2020crossnerevaluatingcrossdomainnamed} addresses domain adaptation with a fully-labeled benchmark spanning five domains, yet, like its predecessors, operates on clean text rather than raw HTML. Tan et al. confirmed the cost of this simplification with HtmlRAG~\cite{Tan_2025}, showing that plain-text conversion strips semantic markers such as tables, headings, and hierarchies that are essential for deep document understanding. Existing structured extraction benchmarks therefore remain either too small, too narrow, or disconnected from the HTML source that real-world systems must process.

\paragraph{Multimodal and Generative Approaches.}
The rise of vision-language models has expanded the extraction landscape: the agentic workflows of LangChain~\cite{langchain_benchmarks} and Image2Struct~\cite{roberts2024image2structbenchmarkingstructureextraction} offer new approaches through tool-use orchestration and visual reconstruction, respectively. However, by relying on flat text or screenshots, these methods bypass the underlying DOM tree, missing the structural signals needed for complex live web pages.

Existing resources address individual components of the web extraction pipeline: pre-training corpora provide scale but lack structural ground truth, content extraction benchmarks evaluate HTML-to-text but not schema extraction, structured extraction datasets are small-scale and outdated, and generative benchmarks operate on clean text rather than raw HTML. To the best of our knowledge, ScrapeGraphAI-100k is the first real-usage dataset that exposes the full extraction four-tuple introduced in Section~\ref{sec:intro} at scale, with per-example schema-validation labels (raw HTML is out of scope for v1.0, see Section~\ref{sec:conclusion}). Schema compliance here denotes structural conformance checked by \texttt{jsonschema-rs}\blue{;} it does not imply semantic correctness, and we preserve this distinction throughout the paper.

\section{ScrapeGraphAI-100k}

\subsection{Data Source and Collection}
Data was collected through PostHog~\footnote{\url{https://posthog.com/}}, an analytics platform integrated into the ScrapeGraphAI library \cite{Perini_Scrapegraph-ai_2024}. The raw dataset contains $\sim$9 million telemetry events collected in Q2--Q3 2025\blue{.} Telemetry was enabled by default in the library during this period. Users could opt out through an environment variable, a configuration file entry, or a function call. Events carried a random installation identifier, which we do not release, and we did not link events to user accounts. Page content, prompts, and responses can still contain personal data from the scraped pages, so we audited these fields (Section~\ref{sec:ethics}). This approach allowed us to aggregate real-world usage patterns ranging from simple testing scripts (e.g., price monitoring, real estate scraping, or market research) to complex production pipelines (e.g., full articles with metadata) without relying on synthetic data generation. Each event captures the full extraction lifecycle: the input prompt, the target JSON schema, the source content, the specific LLM used, and the generated response.

\subsection{Data Processing}\label{sec:processing}

The raw telemetry of $\sim$9 million events contained redundancy, noise, and extreme class imbalance. We implemented a multi-stage cleaning pipeline to produce a high-quality dataset suitable for training and evaluation.

\begin{figure}[t]
  \centering
  \includegraphics[width=\textwidth]{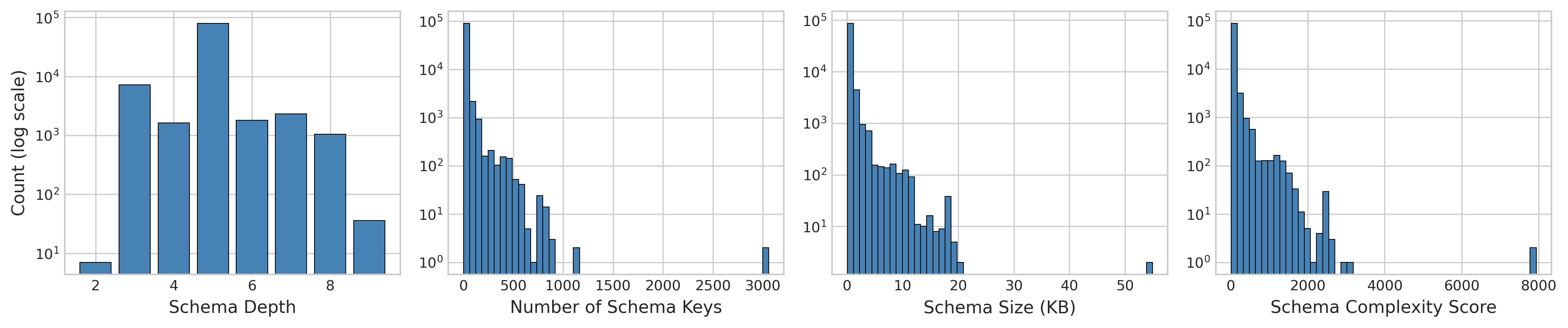}
  \caption{Schema complexity distributions in ScrapeGraphAI-100k (depth, key count, elements, cyclomatic complexity, composite score) across 93,695 extraction events, highlighting a dense core and a long tail of complex schemas.}
  \label{fig:schema-complexity}
\end{figure}

\begin{figure}[t]
  \centering
  \includegraphics[width=\textwidth]{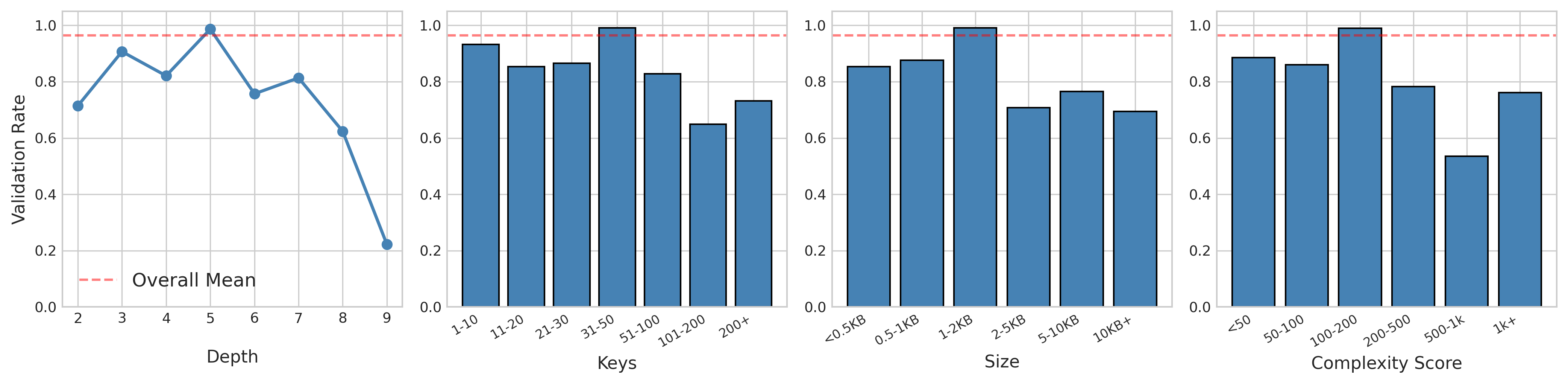}
  \caption{Schema validation rate versus complexity (depth, key count, composite score), the dashed line marks the corpus mean and shows declining validity at higher complexity.}
  \label{fig:validation-complexity}
\end{figure}

\paragraph{Flattening and Validation.}
Each event was flattened into a structure containing the source URL, the user-provided JSON schema, the website content in Markdown, and the LLM-generated response. We validated every response against its corresponding schema using the jsonschema-rs library.\footnote{\url{https://github.com/Stranger6667/jsonschema}} The overall validity rate across the raw corpus was 93\%, in the released dataset it is 96.5\% (90,397 of 93,695). Invalid responses are retained in the final dataset with \texttt{response\_is\_valid=false}, enabling research on extraction failure modes.

\paragraph{Known Noise Modes.}
The 3.5\% of released responses that fail schema validation come from three distinguishable causes: upstream LLM errors (truncation, refusals), fetch-time failures that surfaced in the content field (HTTP error pages, anti-bot challenges, CAPTCHA interstitials, JavaScript-required warnings), and pages whose visible content is primarily non-textual (image galleries, video pages) where Markdown conversion produces little extractable signal. All three are retained in the release with \texttt{response\_is\_valid=false} so that extraction-failure analysis is a first-class use of the dataset. Users intending to train or evaluate on clean extractions should filter on \texttt{response\_is\_valid=true} and apply additional content heuristics for non-textual sources.

\paragraph{Complexity Metrics.}
For each schema, we computed five structural complexity metrics following SLOT~\cite{wang2025slotstructuringoutputlarge}: schema depth, key count, element count, cyclomatic complexity, and a composite complexity score. Cyclomatic complexity is adapted to JSON schemas as $C = E - N + 2P$ over the schema's parse-tree decision graph, with edges $E$, nodes $N$, and connected components $P$, each \texttt{oneOf}/\texttt{anyOf}/\texttt{enum} branch and each conditional \texttt{if}/\texttt{then}/\texttt{else} node counts as an additional decision point. These metrics quantify extraction task difficulty and are detailed in Section~\ref{sec:schema-complexity}.

\paragraph{Balancing and Deduplication.}
The raw dataset exhibited extreme class imbalance: two schemas targeting high-traffic Chinese e-commerce sites accounted for nearly 5 million of the 9 million events. To address this, we grouped events by a SHA256 hash of their JSON schema and applied a sampling strategy retaining at most 5 examples per unique schema. The cap of 5 was chosen to retain the long tail of rare schemas while bounding the contribution of the most frequent ones, trading raw volume for schema diversity at the target corpus size. To ensure diversity, we aimed for selected examples to originate from distinct root domains, which limits how often one website repeats within a schema, although this was not strict: the released rows form only 61,858 unique schema-and-domain pairs. Following these steps, the dataset was reduced from 9 million to 93,695 balanced examples.

\subsection{Dataset Schema, Statistics, and Characteristics}
Each entry in ScrapeGraphAI-100k represents a single extraction event, organized into six field groups: identifiers (id, source, schema\_hash), input content (prompt, schema, content), output (response, response\_is\_valid), execution metadata (llm\_model, execution\_time), size metrics (response\_size, schema\_size), and schema complexity metrics (depth, keys, elements, cyclomatic complexity, composite score). The full field schema and aggregate descriptive statistics are reported in Appendix~\ref{sec:dataset-fields}, Tables~\ref{tab:freq} and~\ref{tab:descriptive_stats}. The main dataset (93,695 examples) is released as a single training set on HuggingFace. A separate, filtered fine-tuning subset with GPT-5-nano-regenerated targets is distributed with a deterministic 25,244/2,808 train/test split produced by random per-row sampling with a fixed seed, plus a 100-example human-verified \texttt{human\_eval} split at tag v1.1(Section~\ref{sec:experiments}).

\paragraph{Schema Complexity}\label{sec:schema-complexity}

Schema complexity in ScrapeGraphAI-100k reflects real-world usage patterns (Table~\ref{tab:descriptive_stats}, Figure~\ref{fig:schema-complexity}). Schema depth clusters tightly around 5 levels (median=5, $\sigma$=0.74), ranging from flat structures (depth 2) to deeply nested hierarchies (depth 9). Key counts show greater variance: the median schema contains 37 keys, but the distribution is heavily right-skewed (max 3,060 keys, $\sigma$=42.10), reflecting the gap between routine extractions (product details, article metadata) and complex multi-entity tasks. The composite complexity score (median=131.3, $\sigma$=123.78) shows two distinct populations. Approximately 90\% of schemas cluster around common extraction templates, while the remaining 10\% reach complexity scores of up to 7,935, representing the challenging edge cases where LLM extraction systems are most likely to fail. This bimodal distribution enables both robust baseline training on routine schemas and stress-testing on structurally demanding tasks.

\paragraph{Content and Response Size}

The \texttt{response\_size} field records the size of the page content, not of the LLM output. Content size is heavy-tailed (Figure \ref{fig:validation-complexity}), with a median of around 10~KB and a maximum near 1~MB. LLM responses are far smaller, with a median of 134 characters and a maximum of 163,805 characters.

\paragraph{Model Distribution and Validation}

GPT-4o-mini dominates the dataset (87.4\% of requests), followed by GPT-4o (3.6\%) and GPT-4.1-mini (1.0\%), with the remainder distributed across Gemini, DeepSeek, Llama, and others. This skewed distribution reflects real-world adoption patterns but represents a limitation of the dataset: extraction patterns and failure modes may be biased toward the behavior of GPT-4o-mini. We observe consistent complexity-driven failure patterns (Figure~\ref{fig:validation-complexity}). Validation rates remain high ($\sim$95\%) for moderate schemas but show sharp, non-linear declines at specific thresholds: depth $\geq 7$ reduces the validation rate from 98\% to 20\%, key count $\geq 200$ reduces it to $\sim$65\%, and complexity score $>1{,}000$ reduces it to $\sim$55\%. These thresholds suggest that LLMs handle incremental complexity gracefully until they reach structural limits, beyond which failures accelerate rapidly.

\paragraph{Domain Coverage}

The dataset spans 80,213 unique URLs from 10,295 hostnames across 8,155 root domains. Coverage is heavy-tailed: the top-10 domains hold 44.2\% of examples, the top-100 hold 69.2\%, and 56.3\% of domains contribute one example. Most of the largest domains are book retailers and academic publishers, which reflects the template described in Section~\ref{sec:processing}, so schema-hash counts overstate task diversity on these domains (Appendix~\ref{sec:domain-diversity}).

\subsection{Comparison with Available Datasets}

Table~\ref{tab:comparison} situates ScrapeGraphAI-100k among related resources. Pre-training corpora such as C4~\cite{raffel2020exploring} and FineWeb~\cite{penedo2024the} offer large scale but discard structural information during preprocessing. Content extraction benchmarks such as WebMainBench~\cite{liu2025drippertokenefficientmainhtml} evaluate HTML-to-text conversion without providing structured output targets. Traditional structured extraction datasets, including SWDE~\cite{hao2011swde}, OpenCeres~\cite{lockard-etal-2019-openceres}, and CrossNER~\cite{liu2020crossnerevaluatingcrossdomainnamed}, are restricted to English, operate on raw HTML, and predate LLM-based extraction methods. Schema benchmarks such as JSONSchemaBench~\cite{jsonschemabench2025} assess structured generation using synthetic prompts that lack web content, LangChain~\cite{langchain_benchmarks} operates on clean text rather than realistic web inputs. MS MARCO~\cite{nguyen2016msmarco} targets free-form QA rather than schema-constrained extraction. ScrapeGraphAI-100k addresses these gaps by exposing the full four-tuple at scale across 15 fastText-identified named languages plus a long tail aggregated as \emph{Other}. Content is predominantly English, at 88.9\% of detected content (Appendix~\ref{sec:content-language}).

\begin{table}[th!]
\caption{Comparison of ScrapeGraphAI-100k with existing datasets.}
\label{tab:comparison}
\footnotesize
\centering
\begin{tabularx}{0.9\linewidth}{X r l l l}
\toprule
\textbf{Dataset} & \textbf{Size} & \textbf{Lang.} & \textbf{Input} & \textbf{Output} \\
\midrule
C4 & 172B & English & Text &  \\
FineWeb & 15T & English & Text &  \\
\midrule
WebMainBench & 7,887 & English & HTML & Markdown \\
WCE Benchmark & $\sim$7K & English & HTML & Text \\
\midrule
SWDE & 124k & English & HTML & KV \\
OpenCeres & 2M & English & HTML & Triples \\
CrossNER & $\sim$2.5k & English & Text & Entities \\
\midrule
JSONSchema & 10K &  & JSON & JSON \\
LangChain & N/A & English & Text & Struct. \\
\midrule
MS MARCO & 8.8M & English & Text & Answers \\
\midrule
\textbf{ScrapeGraphAI-100k} & \textbf{93k} & \textbf{Multilingual} & \textbf{Markdown} & \textbf{JSON} \\
\bottomrule
\end{tabularx}
\end{table}

\section{Utility and Research Opportunities}\label{sec:utility}

ScrapeGraphAI-100k supports four research directions that leverage its scale, structural metadata, and real-world provenance. (i) Efficient extraction with small language models. The 93,695 examples and 90/10 split between routine and edge-case schemas enable PEFT-style fine-tuning~\cite{xu2023parameterefficientfinetuningmethodspretrained} of compact open models (e.g.\ Qwen3-1.7B, Llama-3.1-8B) to approach proprietary baselines, supporting privacy-preserving on-premises extraction. (ii) Stratified evaluation of extraction systems. The depth, key-count, and cyclomatic-complexity fields enable analysis beyond aggregate accuracy, exposing thresholds (e.g.\ depth $\geq 7$, keys $\geq 200$, Section~\ref{sec:schema-complexity}) usable for adaptive task routing~\cite{ong2025routellmlearningroutellms}, and the \texttt{human\_eval} split supports correctness-level evaluation at small scale. (iii)~ Failure-mode analysis. Retained invalid responses (\texttt{response\_is\_valid=false}) and their correlation with complexity metrics (Figure~\ref{fig:validation-complexity}) enable systematic study of syntactic, structural, and semantic failures and pre-execution failure prediction~\cite{kadavath2022languagemodelsmostlyknow, manakul2023selfcheckgptzeroresourceblackboxhallucination}. (iv)~ Schema induction. The 25,453 schema hashes paired with the content that produced them provide training signal for ``schema-from-content''~\cite{tang2023harvestingeventschemaslarge}, bootstrapping extraction pipelines for new domains without manual schema design.

\section{Availability and Maintenance}\label{sec:availability}

We release three artifacts, cross-linked from the project README, all under Apache 2.0: the main dataset (scrapegraphai/scrapegraphai-100k\footnote{\url{https://huggingface.co/datasets/scrapegraphai/scrapegraphai-100k}}, 93,695 examples) carrying Markdown content, prompts, JSON schemas, LLM responses, structural-validation flags, schema-complexity metrics, and (domain-masked) source URLs, the fine-tuning subset\footnote{\url{https://huggingface.co/datasets/scrapegraphai/scrapegraph-100k-finetuning}} of Section~\ref{sec:experiments} with GPT-5-nano-regenerated targets, and the fine-tuned Qwen3-1.7B model\footnote{\blue{\url{https://huggingface.co/scrapegraphai/sgai-qwen3-1.7b}}}. Reproducibility code is at \url{https://github.com/ScrapeGraphAI/scrapegraph-100k-paper}, and the dataset loads via load\_dataset("scrapegraphai/scrapegraphai-100k"). v1.0 does not redistribute raw HTML/WARC sources, which are scheduled for v2.0 (Section~\ref{sec:conclusion}). Results in this paper use tag v1.0 of the main dataset (commit \texttt{0e6ad2d}) and tag v1.0 of the fine-tuning subset (commit \texttt{2c2371f}). Tag v1.1 of the fine-tuning subset adds the \texttt{human\_eval} split. We adopt semantic versioning, plan a longitudinal refresh \blue{from Q2--Q3 2026 telemetry in 2027}, and accept community contributions via GitHub issues and pull requests.

\section{Case Study: Fine-Tuning a Small Language Model}\label{sec:experiments}

To illustrate how ScrapeGraphAI-100k can be used, we present a case study in knowledge distillation. This is not a systematic benchmark, it demonstrates one concrete way the dataset supports downstream research. We show that a small language model (Qwen3-1.7B \cite{qwen3technicalreport}) can approach the teacher agreement of significantly larger models when fine-tuned on teacher-generated labels derived from our dataset, suggesting practical utility for cost-effective, privacy-preserving, on-premises extraction.

\paragraph{Experimental Setup}
We curate a training-efficient subset from ScrapeGraphAI-100k, filtering for schema length ($\le$10k chars), source content ($\le$50k chars), and response size ($\le$10k chars). We then performed a re-generation pass using GPT-5-nano to produce schema-consistent target outputs. These teacher outputs are pseudo-labels, not human-verified ground truth, we train the student to imitate them, so the student's metrics measure agreement with the teacher on structural accuracy rather than absolute correctness. Schema compliance rises from 83.7\% to 100\%, because outputs that fail validation are discarded. Regeneration largely preserves structure (flattened-key Jaccard 0.80), but 74.4\% of shared leaf values change (Appendix~\ref{sec:regeneration}). Large content fields were processed using a 4,096-token sliding window to maintain context without exceeding the model's primary attention window. This version of dataset is published via Huggingface\footnote{\url{https://huggingface.co/datasets/scrapegraphai/scrapegraph-100k-finetuning}} for further research.

\paragraph{Training Methodology}
Fine-tuning was conducted using QLoRA~\cite{dettmers2023qlora} with 4-bit quantization. We applied LoRA \cite{hu2021loralowrankadaptationlarge} adapters (rank $r=16$) to all linear projections within the attention and MLP blocks. We utilized a learning rate of $10^{-4}$ with a cosine decay scheduler over 2 epochs on a single NVIDIA A100 (80GB). We specifically contrast standard fine-tuning against completion-only loss (response-only training), where the gradient is computed exclusively on the assistant's output tokens to prioritize structural accuracy in the JSON response.

\paragraph{Evaluation}

We evaluate extraction quality across complementary dimensions using vLLM~\cite{kwon2023efficient} with greedy decoding. Structural validity metrics assess whether outputs are well-formed: is\_valid\_json checks JSON parseability, while is\_schema\_compliant verifies conformance to the target schema. Key extraction metrics (precision, recall, F1) measure structural accuracy by comparing flattened JSON keys using dot-notation paths with [*] wildcards for arrays. Value extraction is captured by value\_score, a type-aware metric averaging exact match for booleans/numbers, set equality for arrays, and sentence-level BLEU~\cite{papineni-etal-2002-bleu} for strings.\footnote{Sentence-level BLEU is known to degrade on very short strings (1--5 tokens), which is common for extracted field values like titles, prices, and names, we use it as one signal of lexical overlap rather than a definitive value-correctness measure, and discuss a complementary exact-match metric in Appendix~\ref{sec:value-score-alt}.} Structural-accuracy claims in the paper rest on key-based metrics and schema compliance, which are unaffected by this caveat. Finally, overall\_bleu on serialized JSON provides a holistic quality measure. Evaluation uses the test split of the filtered fine-tuning subset (2,808 examples), samples exceeding the 8192-token context limit are excluded, leaving 2,714. Targets in this split are GPT-5-nano-regenerated references rather than original user-facing telemetry responses, so reported metrics measure student/teacher agreement rather than agreement with real production outputs.

Both the Qwen3 baselines and the fine-tuned model use greedy decoding under vLLM with the following prompt template. We deliberately do not apply schema-constrained decoding, JSON-mode, or few-shot exemplars to either side, so the comparison reflects raw generation quality under a uniform inference recipe.

\begin{lstlisting}[language=Python, basicstyle=\ttfamily\scriptsize, breaklines=true, frame=single, columns=fullflexible, keepspaces=true]
Extract data from the content according to the JSON schema.
Schema: {schema}
Content: {content}
Return ONLY valid JSON matching the schema.
\end{lstlisting}

\begin{figure}[t]
 \centering
 \includegraphics[width=0.8\linewidth]{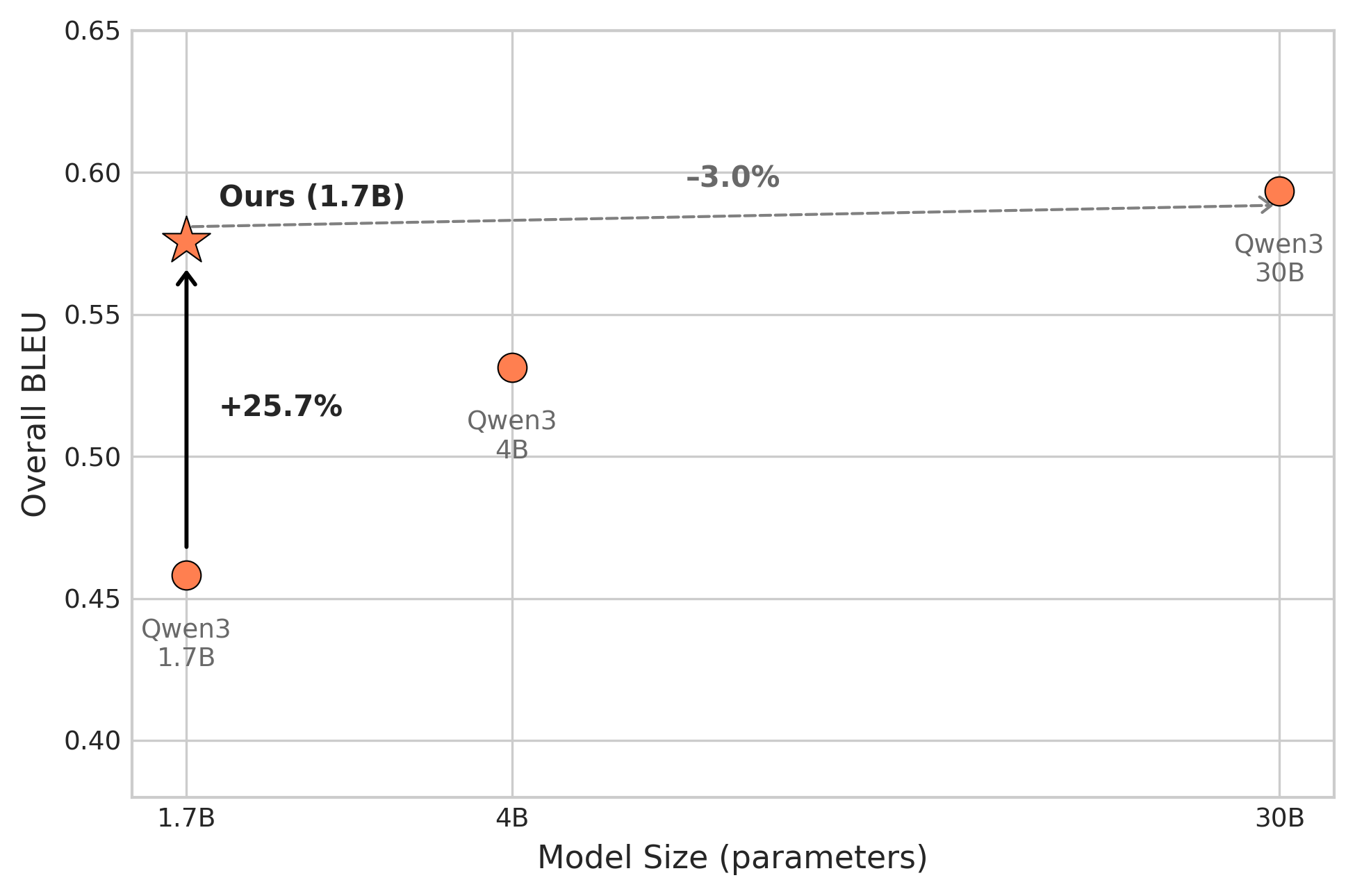}
 \caption{Overall BLEU vs. model size on the evaluation set. Circles are Qwen3 baselines (1.7B/4B/30B-A3B) and the star is our fine-tuned 1.7B model, which gains +25.7\% over the 1.7B baseline and comes within 3.0\% of the 30B-A3B (3.3B active) MoE model. All scores are agreement with GPT-5-nano targets.}
 \label{fig:model_scaling}
\end{figure}

\paragraph{Results}
Table~\ref{tab:finetune-results} reports per-metric results, and Figure~\ref{fig:model_scaling} visualizes the overall BLEU scaling trend. The 30B-A3B (3.3B active) reference model achieves the strongest validity and content scores (e.g., 0.9913 valid JSON, 0.8915 key F1). Our fine-tuned 1.7B stays close on structural accuracy (key F1 0.8866, within 0.005 of 30B-A3B) and surpasses the 4B baseline on all key-based metrics, showing that ScrapeGraphAI-100k can narrow the gap between small and large models on agreement with GPT-5-nano targets.

\paragraph{Discussion} The comparison is asymmetric: ``Ours'' is fine-tuned to imitate GPT-5-nano and then evaluated against GPT-5-nano-regenerated targets, whereas the Qwen3 baselines have no exposure to that teacher. The reported gaps therefore conflate baseline-to-teacher distribution distance with capability difference and likely overstate the absolute gap to the baselines, a symmetric comparison would fine-tune the baselines on the same teacher targets.

Response-only training improves over the base 1.7B across all metrics. The largest gain is in schema compliance, rising from 0.7954 to 0.9167 (a relative increase of 15\%), suggesting that optimising for output generation helps models internalise schema constraints. Extra key generation drops from 2.64 to 0.33 per sample and even undercuts the 30B-A3B baseline (0.5679), indicating that dataset-specific fine-tuning reduces hallucinated fields beyond what scale alone provides. A notable gap remains in value extraction: the fine-tuned model reaches 89\% key F1 but only 0.4624 on value score, showing that identifying which fields to extract is easier than extracting what they contain. This structural-semantic gap, where models produce correct JSON skeletons with imprecise values, is an open challenge for extraction systems, although its precise magnitude depends on the metric since the string component of \texttt{value\_score} inherits sentence-level BLEU's known weakness on short strings (Appendix~\ref{sec:value-score-alt}).

 \begin{table}[t]
 \centering
 \caption{Agreement with GPT-5-nano targets on the fine-tuning subset test split (N=2,714 after context-length filtering). Best results per metric in \textbf{bold}. $\downarrow$ indicates lower is better.}
 \label{tab:finetune-results}
 \small
 \begin{tabular}{@{}lcccc@{}}
 \toprule
 \textbf{Metric} & \textbf{Qwen3 1.7B} & \textbf{Qwen3 4B} & \shortstack{\textbf{Qwen3-30B-A3B}\\\textbf{(3.3B active)}} & \textbf{Ours} \\
 \midrule
 \multicolumn{5}{@{}c}{\textit{Validity Metrics}} \\[2pt]
 Valid JSON & 0.9544 & 0.9681 & \textbf{0.9913} & 0.9558 \\
 Schema compliant & 0.7954 & 0.8301 & \textbf{0.9363} & 0.9167 \\
 \midrule
 \multicolumn{5}{@{}c}{\textit{Key-based Metrics}} \\[2pt]
 Key precision & 0.7332 & 0.8029 & \textbf{0.8997} & 0.8988 \\
 Key recall & 0.7477 & 0.8069 & \textbf{0.8956} & 0.8864 \\
 Key F1 & 0.7343 & 0.7995 & \textbf{0.8915} & 0.8866 \\
 Missing keys $\downarrow$ & 1.8997 & 1.1706 & \textbf{0.8156} & 1.0177 \\
 Extra keys $\downarrow$ & 2.6353 & 1.0659 & 0.5679 & \textbf{0.3300} \\
 \midrule
 \multicolumn{5}{@{}c}{\textit{Content Metrics}} \\[2pt]
 Value score & 0.2735 & 0.3655 & \textbf{0.4797} & 0.4624 \\
 Overall BLEU & 0.4581 & 0.5314 & \textbf{0.5935} & 0.5759 \\
 \bottomrule
 \end{tabular}
 \end{table}

\paragraph{Human-Verified Evaluation}

Teacher targets are not ground truth, so we built \texttt{human\_eval}: 100 test examples, stratified by schema complexity and page truncation, leak-free with respect to training, with every leaf value verified or corrected by three of the authors against the source page content (Appendix~\ref{sec:human-protocol}). Annotators edited the teacher draft in 77 of 100 examples, changing 272 of 858 leaf values (31.7\%), adding 23, and removing 44. Table~\ref{tab:human-eval} shows that structural metrics overstate correctness: the teacher is schema-compliant by construction, yet about a third of its leaf values needed correction. Two caveats apply. Gold starts from teacher drafts, which anchors it toward the teacher and favors the teacher in any comparison. Value score uses sentence-level BLEU for strings (Appendix~\ref{sec:value-score-alt}), so it measures similarity to gold and is not a share of correct values.

 \begin{table}[t]
 \centering
 \caption{Scores against human-verified gold on \texttt{human\_eval} (n=100), with 95\% bootstrap confidence intervals. Teacher compliance is 1.000 by construction. Frontier rows come from separate 100-row runs (Appendix~\ref{sec:frontier}).}
 \label{tab:human-eval}
 \small
 \begin{tabular}{@{}lccc@{}}
 \toprule
 \textbf{Model} & \textbf{Schema compliant} & \textbf{Key F1} & \textbf{Value score} \\
 \midrule
 GPT-5-nano (teacher) & 1.000 & 0.963 [0.929, 0.989] & 0.640 [0.573, 0.710] \\
 GPT-5.6 & 1.000 & 0.986 [0.973, 0.996] & 0.595 [0.529, 0.663] \\
 Claude Sonnet 4.6 & 0.980 [0.950, 1.000] & 0.940 [0.898, 0.975] & 0.499 [0.433, 0.568] \\
 Ours (1.7B) & 0.900 [0.840, 0.950] & 0.887 [0.831, 0.937] & 0.452 [0.382, 0.525] \\
 \bottomrule
 \end{tabular}
 \end{table}

\paragraph{Frontier Baselines}

To see whether complexity failures reach production systems, we ran GPT-5.6 and Claude Sonnet 4.6 with native structured-output modes on the test split (N=2,761). Over all calls, compliance is 0.988 and 0.983 and key F1 is 0.929 and 0.915, against 0.919 and 0.887 for our model, and key F1 falls with schema complexity for all three (Appendix~\ref{sec:frontier}).

\section{Ethical Considerations}\label{sec:ethics}

ScrapeGraphAI-100k is built from telemetry that was enabled by default in the ScrapeGraphAI library. Users could opt out, and the library documented how, but they were not asked to agree to public release of their events, so this is weaker than opt-in consent. The dataset is released under Apache 2.0, and the license covers the derivative artifacts we release (Markdown content, prompts, schemas, responses, metadata), not the underlying web pages, whose copyright remains with their original publishers. We deliberately do not redistribute raw HTML or exhaustive page reproductions. The corpus exhibits two biases researchers should consider: GPT-4o-mini accounts for 87.4\% of requests, and book retail and academic publisher pages remain over-represented because of one dominant template (Section~\ref{sec:processing}). ScrapeGraphAI is a general-purpose tool, compliance with source-website Terms of Service, robots.txt, and applicable laws is the responsibility of individual users, we cannot independently verify that all scraped content was obtained accordingly, and we do not endorse scraping that violates such policies. Finally, the GPT-5-nano targets used in the fine-tuning subset (Section~\ref{sec:experiments}) are schema-valid pseudo-labels, not human-verified ground truth.

\paragraph{PII audit and takedown.}
We scanned the \texttt{prompt}, \texttt{content}, \texttt{schema}, and \texttt{response} fields of all 93,695 rows with openai/privacy-filter and eight credential-format patterns (Appendix~\ref{sec:pii-audit}). The most frequent detections are person names (43.3\% of rows) and URLs (35.7\%), and emails, phone numbers, and addresses each appear in 4--6\% of rows. Detections concentrate in \texttt{content} and typically cover names, bylines, and business contact details on public pages, which we retain as benchmark ground truth. Credential-format strings appear in at least 52 rows, and we did not test whether any are valid. These rows remain in v1.0 and are removable through the takedown process. Takedown requests go to \texttt{support@scrapegraphai.com}, are acknowledged within 24 hours, and, if accepted, are applied in a patch release within 30 days and propagated to the fine-tuning subset.

\section{Limitations}\label{sec:limitations}
The release supports three evaluative uses: (i) measuring how structural-compliance rates degrade as JSON-schema complexity grows under realistic input distributions, (ii) serving as training data for knowledge-distillation of small language models into schema-constrained extractors, and (iii) enabling stratified analysis of extraction behavior across schema-structural dimensions (depth, key count, cyclomatic complexity). These claims rest on three scoping assumptions: the GPT-5-nano teacher used in the fine-tuning subset is a reference signal rather than ground truth, so case-study numbers measure student/teacher agreement, the Markdown surface is the unit of content we characterize, with raw HTML and DOM out of scope for v1.0, ScrapeGraphAI-telemetry population is a proxy for typical LLM-based extraction workloads, skewed toward one book-contents template and GPT-4o-mini. Correspondingly, we do \emph{not} claim semantic correctness of individual extractions beyond the 100-example \texttt{human\_eval} split, and do \emph{not} evaluate on HTML or DOM inputs. ScrapeGraphAI-100k v1.0 has eight principal limitations expanded in Appendix~\ref{sec:limitations-detail}.

\section{Conclusion}\label{sec:conclusion}
ScrapeGraphAI-100k is a real-world dataset for LLM-based structured web extraction, derived from 9M production telemetry events into 93,695 schema-balanced examples that pair Markdown content with natural-language prompts, JSON schemas, schema-validated responses, and fine-grained complexity metadata. Our analysis exposes sharp, non-linear failure thresholds as schema depth and key count increase, and the case study shows that fine-tuning a 1.7B model on this data narrows the gap to a 30B-A3B (3.3B active) MoE baseline on structural accuracy against teacher targets, while the \texttt{human\_eval} split shows how far those targets are from correct values. v2.0 \blue{(planned for 2027)} will add a raw-HTML/WARC companion, a larger human-validated benchmark with inter-annotator agreement, per-row LLM-judge quality tiers, explicit failure-mode labels, and broader multilingual and model coverage. In a 1,000-example pilot of that judge, reweighted to the dataset distribution, 53.4\% of examples reach the top tier, 19.4\% the second, 15.8\% the third, and 11.4\% are rejected. The dataset is publicly available on HuggingFace to enable immediate research and reproducible comparisons.

\section{Broader Impact}\label{sec:broader-impact}

ScrapeGraphAI-100k supports research on schema-constrained generation, structured extraction, and small-model distillation for privacy-preserving deployment. Because the corpus could also fine-tune extraction agents for automated scraping, users should weigh dual-use implications and respect Terms of Service, robots.txt, and applicable laws (Section~\ref{sec:ethics}). That said, the data comes from telemetry of a public open-source library (Section~\ref{sec:processing}), so the uplift over running the library directly is modest. The dataset is released on Hugging Face under Apache 2.0 with the takedown process in Section~\ref{sec:ethics}.

\begin{ack}
ScrapeGraphAI funded this work, including staff time, API usage, and compute. The work received no third-party funding.

\paragraph{Competing interests.} Francesco Zuppichini, Marco Vinciguerra, and Lorenzo Padoan are co-founders or employees of ScrapeGraphAI, and some of them hold equity in the company. The dataset comes from telemetry of the open-source ScrapeGraphAI library, which ScrapeGraphAI maintains. William Brach has no financial relationship with ScrapeGraphAI.
\end{ack}

\bibliographystyle{plainnat}
\bibliography{neurips}

\appendix

\section{Detailed Limitations}\label{sec:limitations-detail}

We expand the eight principal limitations summarized in Section~\ref{sec:limitations}. Each item points to the relevant sections of the main paper instead of repeating their details.

\paragraph{Structural, not semantic, validation.}
The \texttt{response\_is\_valid} label records whether a response parses as JSON and conforms to its schema under \texttt{jsonschema-rs} (Section~\ref{sec:processing}). It does not record whether the extracted values faithfully represent the source page. A structurally valid response may still be factually wrong. The value-score gap in Table~\ref{tab:finetune-results} (0.46 on the fine-tuned model) shows the size of this structural--semantic gap in practice.

\paragraph{\blue{Small human-validated benchmark.}}
\blue{The \texttt{human\_eval} split has 100 examples, reports no inter-annotator agreement, and has gold that starts from teacher drafts (Appendix~\ref{sec:human-protocol}).} The case-study extraction metrics compare student outputs against GPT-5-nano-regenerated pseudo-labels. They therefore measure teacher/student agreement, not correctness (Section~\ref{sec:experiments}). We plan \blue{a larger stratified benchmark} with inter-annotator agreement for v2.0 (Section~\ref{sec:conclusion}). We did not train the baseline models on the GPT-5-nano targets, so the case study may overstate the absolute gap to baselines\blue{.} A fairer comparison would fine-tune baselines on the same targets.

\paragraph{Markdown-only surface.}
The release distributes Markdown-converted derivatives of page content\blue{.} We defer raw HTML and WARC sources to v2.0. Users cannot reproduce research on HTML-specific structural signals (tables, headings, DOM hierarchy) end-to-end from v1.0.

\paragraph{Distribution skew.}
GPT-4o-mini accounts for 87.4\% of requests\blue{. One book-contents template accounts for at least half of rows (Section~\ref{sec:processing}). The top-10 domains hold 44.2\% of examples (Appendix~\ref{sec:domain-diversity}).} Results may generalize imperfectly to other model families or under-represented domains. Schema depth also clusters tightly around 5 levels (Section~\ref{sec:schema-complexity}), so we observe extreme-depth behavior only in the tail.

\paragraph{Temporal snapshot.}
The release covers Q2--Q3 2025 telemetry. Web content evolves. Some source URLs may no longer resolve, or their pages may have changed substantively since we collected them. We plan a refresh \blue{from Q2--Q3 2026 telemetry in 2027 (Section~\ref{sec:availability})}.

\paragraph{\blue{Telemetry population bias.}}
\blue{The data comes from ScrapeGraphAI users who did not opt out of telemetry. These users} may skew toward open-source-library adopters and may not represent all LLM-based extraction workloads.

\paragraph{Known noise modes.}
\blue{Approximately 3.5\% of released responses} fail schema validation (Section~\ref{sec:processing}). Upstream LLM errors, fetch-time failures that appear in the content field, and non-textual pages drive these failures. We retain these rows with \texttt{response\_is\_valid=false}\blue{.} Users who need clean evaluation data should filter on this flag and apply additional content heuristics.

\paragraph{Case-study scope.}
The fine-tuning case study (Section~\ref{sec:experiments}) uses a single model family (Qwen3) and reports single-seed results without error bars. We frame it as a demonstration of dataset utility, not as a systematic benchmark. \blue{Its test split overlaps with training (Section~\ref{sec:experiments}), and only the \texttt{human\_eval} split is leak-free. Its schema mix also differs from the main dataset. No single schema shape exceeds 1\% of the 2,761 evaluated test rows, so the book-contents template does not dominate the test split.}

\section{Dataset Fields and Descriptive Statistics}
\label{sec:dataset-fields}

Tables~\ref{tab:freq} and~\ref{tab:descriptive_stats} list the full field schema and aggregate descriptive statistics for the 93,695 examples in ScrapeGraphAI-100k v1.0.

\begin{table}[h!]
 \caption{Dataset Feature Summary}
 \label{tab:freq}
 \small
 \centering
 \begin{tabular}{@{}lp{8cm}@{}}
  \toprule
  Feature Name & Description\\
  \midrule
  \multicolumn{2}{l}{\textbf{Identifiers}} \\
  \texttt{id} & Unique identifier for each extraction event \\
  \texttt{source} & Source URL (domain-masked if sensitive) \\
  \texttt{schema\_hash} & SHA256 hash for deduplication \\
  \midrule
  \multicolumn{2}{l}{\textbf{Input Content}} \\
  \texttt{prompt} & Natural language instruction \\
  \texttt{schema} & JSON schema defining expected output \\
  \texttt{content} & Cleaned Markdown website content \\
  \midrule
  \multicolumn{2}{l}{\textbf{Output}} \\
  \texttt{response} & The LLM's returned output \\
  \texttt{response\_is\_valid} & True iff the response parsed as JSON and conformed to the schema (structural, not semantic, validation) \\
  \midrule
  \multicolumn{2}{l}{\textbf{Execution Metadata}} \\
  \texttt{llm\_model} & LLM identifier used for extraction \\
  \texttt{execution\_time} & Latency in seconds \\
  \midrule
  \multicolumn{2}{l}{\textbf{Size Metrics}} \\
  \texttt{response\_size} & \blue{Page content size in bytes (despite its name, not the size of the LLM response)} \\
  \texttt{schema\_size} & Schema size in bytes \\
  \midrule
  \multicolumn{2}{l}{\textbf{Complexity Metrics}} \\
  \texttt{schema\_depth} & Maximum nesting depth \\
  \texttt{schema\_keys} & Number of keys in schema \\
  \texttt{schema\_elements} & Total number of schema elements \\
  \texttt{schema\_cyclomatic\_complexity} & Cyclomatic complexity of schema \\
  \texttt{schema\_complexity\_score} & Composite complexity metric \\
  \bottomrule
 \end{tabular}
\end{table}

\begin{table}[h!]
  \caption{Descriptive Statistics of ScrapeGraphAI-100k Dataset (N=93,695)}
  \label{tab:descriptive_stats}
  \small
  \centering
  \begin{tabular}{@{}lrrrrrr@{}}
   \toprule
   Metric & Mean & Std & Median & P25 & P75 & Max \\
   \midrule
   Exec. Time (s) & 14.85 & 75.29 & 8.49 & 4.86 & 15.19 & 14,345 \\
   \blue{Content Size (B)} & 19,772 & 48,476 & 10,007 & 790 & 21,687 & 991,323 \\
   Schema Size (B) & 1,161 & 1,012 & 1,080 & 1,058 & 1,106 & 56,254 \\
   Schema Depth & 4.93 & 0.74 & 5.0 & 5.0 & 5.0 & 9.0 \\
   Schema Keys & 41.40 & 42.10 & 37.0 & 37.0 & 37.0 & 3,060 \\
   Schema Elements & 47.95 & 55.24 & 41.0 & 41.0 & 41.0 & 3,769 \\
   Cyclomatic Compl. & 43.25 & 47.23 & 38.0 & 38.0 & 38.0 & 3,355 \\
   Complexity Score & 141.53 & 123.78 & 131.3 & 131.3 & 131.3 & 7,935 \\
   \bottomrule
  \end{tabular}
 \end{table}

\newpage
\clearpage
\section{Example Dataset Entry}
\label{sec:example-entry}

The following JSON object illustrates a complete entry from ScrapeGraphAI-100k. This example shows a table-of-contents extraction task from a publisher chapter page. We truncate the \texttt{content} field for readability.

\begin{lstlisting}[
 basicstyle=\ttfamily\scriptsize,
 breaklines=true,
 frame=single,
 columns=fullflexible,
 keepspaces=true
]
{
 "schema": "{\"$defs\":{\"ScrapedChapter\":{\"description\":\"Schema of a chapter present in the table of contents.\",\"properties\":{\"title\":{\"description\":\"Title of the chapter without the prefix.\",\"title\":\"Title\",\"type\":\"string\"},\"prefix\":{\"description\":\"Prefix of the chapter (e.g. Part, Chapter, Chapter 2.3)\",\"title\":\"Prefix\",\"type\":\"string\"},\"category\":{\"description\":\"Category of the chapter ('content' or 'other')\",\"title\":\"Category\",\"type\":\"string\"}},\"required\":[\"title\",\"prefix\",\"category\"],\"title\":\"ScrapedChapter\",\"type\":\"object\"}},\"description\":\"Output schema of the table of contents scraper.\",\"properties\":{\"certainty_percentage\":{\"default\":0,\"description\":\"Certainty that the scraped chapters belong to the book named The Routledge international companion to multicultural education published by Taylor & Francis and has edition 1st\",\"maximum\":100,\"minimum\":0,\"title\":\"Certainty Percentage\",\"type\":\"integer\"},\"chapters\":{\"default\":[],\"description\":\"Chapters found in the table of contents\",\"items\":{\"$ref\":\"#/$defs/ScrapedChapter\"},\"title\":\"Chapters\",\"type\":\"array\"}},\"title\":\"ScraperOutput\",\"type\":\"object\"}",
 "content": "[Skip to main content](/chapters/edit/10.4324/9780203881514-11/multicultural-education-united-kingdom-sally-tomlinson#main-content) ... [truncated for brevity]",
 "response": "{\"certainty_percentage\":100,\"chapters\":[{\"title\":\"Multicultural education in the United Kingdom\",\"prefix\":\"Chapter\",\"category\":\"content\"}]}"
}
\end{lstlisting}

\paragraph{Field Annotations.}
The \texttt{schema} field defines the expected JSON structure in JSON Schema syntax. The dataset stores it as a string. The \texttt{content} field holds the cleaned Markdown version of the source webpage (truncated here). The \texttt{response} field holds the LLM's extracted output, also as a JSON string.

\section{Metric Correlations}
\label{sec:metric-correlations}

\begin{figure}[h]
  \centering
  \includegraphics[width=0.8\linewidth]{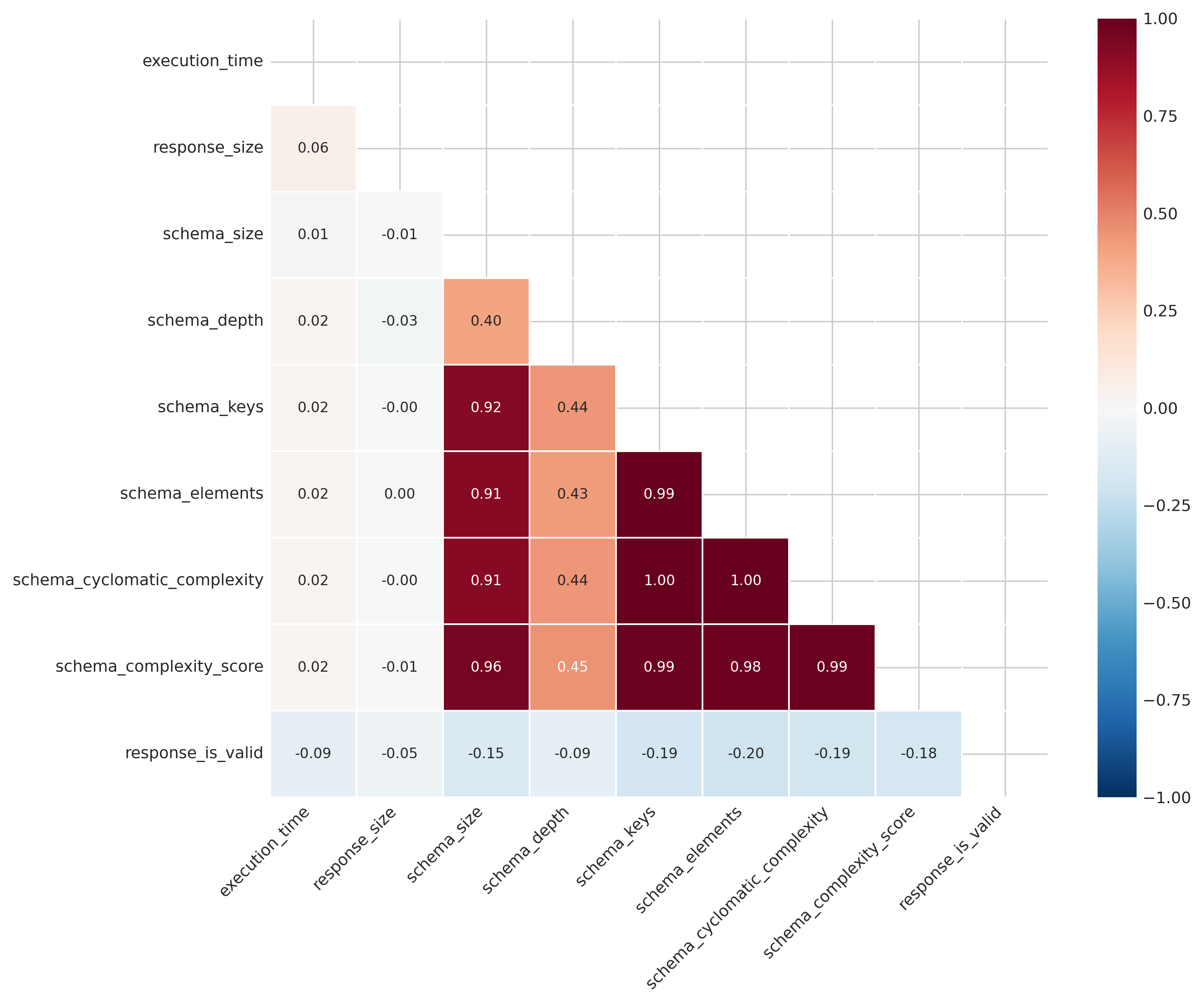}
  \caption{Correlation matrix of dataset metrics, showing strong correlations among complexity measures and a negative relationship between complexity and validation. \blue{The \texttt{response\_size} variable records page content size.}}
  \label{fig:correlation_matrix}
\end{figure}

The correlation structure of dataset metrics (Figure~\ref{fig:correlation_matrix}) reveals three findings with practical implications for extraction system design.
\begin{enumerate}
 \item Schema complexity metrics show near-perfect multicollinearity. Keys, elements, cyclomatic complexity, and composite score correlate at $r \geq 0.98$. This indicates that key count alone suffices as a complexity proxy for most applications. For downstream stratification, we recommend reporting key count as the primary complexity axis and treating element count, cyclomatic complexity, and composite score as redundant. Depth correlates only moderately with them ($r = 0.43$--$0.45$) and is worth reporting alongside key count.
 \item Validation success correlates negatively with every complexity measure, from $r = -0.09$ for depth to $r = -0.20$ for element count. All these correlations are negative but weak, so schema complexity predicts extraction failure risk consistently but only weakly. Adaptive prompting or model selection can exploit this pattern.
 \item Execution time correlates negligibly with schema complexity ($r \approx 0.02$) and only weakly with \blue{page content size (the \texttt{response\_size} field)} ($r = 0.06$). This suggests that factors outside schema structure \blue{and content size} (e.g., API overhead, model inference) dominate latency in LLM-based extraction.
\end{enumerate}
These patterns let practitioners use lightweight complexity estimates for task routing and focus optimization on non-schema bottlenecks.

\section{Complementary Metric for String Values}
\label{sec:value-score-alt}

The \texttt{value\_score} component for string fields uses sentence-level BLEU (Section~\ref{sec:experiments}). Extracted field values are often short, typically 1--5 tokens (titles, prices, names, identifiers). BLEU's brevity penalty and n-gram coverage degrade sharply at these lengths. A single BLEU number can therefore over-credit lexical overlap, because BLEU-1 dominates when higher-order n-grams are absent. It can also under-credit minor surface variation, such as alternate spellings, capitalization, or whitespace.

We recommend reporting \emph{normalized exact-match} as a complementary metric on the same string-typed fields. It has four steps:
\begin{enumerate}
 \item Lowercase the prediction and the reference.
 \item Collapse internal whitespace.
 \item Strip surrounding punctuation.
 \item Count a match only if the two strings are equal.
\end{enumerate}
Exact-match is the natural floor on string-value correctness and is robust to length. However, it is conservative in the opposite direction: close-but-not-identical strings score zero. Together, BLEU and exact-match bracket value-level performance. BLEU acts as a rough upper bound from lexical overlap, and exact-match acts as the strict floor.

\section{Content Language Distribution}
\label{sec:content-language}

We identified the language of each \texttt{content} field with the fastText language identification model (facebook/fasttext-language-identification)\blue{. Before detection, we stripped raw URLs and the Markdown syntax for links and images, because detection on raw markup misattributes URL-heavy pages to CJK languages. This leaves 87,887 examples with enough visible text. The other 5,808 were too short or undetectable.} fastText reliably detects \emph{script} but cannot distinguish Cantonese from Standard Written Chinese in Traditional script\blue{. We therefore} report the Traditional-script Chinese share as \texttt{yue\_Hant} (Traditional Chinese), not as Cantonese. \blue{Table~\ref{tab:content-language-stats} reports the 15 most frequent languages plus a long tail aggregated as \emph{Other}.}

\begin{table}[h]
 \caption{\blue{Top content languages in ScrapeGraphAI-100k after stripping markup and URLs (N=87,887).}}
 \label{tab:content-language-stats}
 \small
 \centering
 \blue{\begin{tabular}{@{}lrr@{}}
  \toprule
  \textbf{Language} & \textbf{Count} & \textbf{Percentage (\%)} \\
  \midrule
  \texttt{eng\_Latn} & 78,088 & 88.85 \\
  \texttt{yue\_Hant} & 2,233 & 2.54 \\
  \texttt{kor\_Hang} & 2,027 & 2.31 \\
  \texttt{deu\_Latn} & 1,406 & 1.60 \\
  \texttt{nld\_Latn} & 695 & 0.79 \\
  \texttt{spa\_Latn} & 505 & 0.57 \\
  \texttt{ita\_Latn} & 428 & 0.49 \\
  \texttt{dan\_Latn} & 344 & 0.39 \\
  \texttt{fur\_Latn} & 324 & 0.37 \\
  \texttt{swe\_Latn} & 312 & 0.36 \\
  \texttt{fra\_Latn} & 311 & 0.35 \\
  \texttt{por\_Latn} & 185 & 0.21 \\
  \texttt{cat\_Latn} & 147 & 0.17 \\
  \texttt{ind\_Latn} & 113 & 0.13 \\
  \texttt{rus\_Cyrl} & 110 & 0.13 \\
  \addlinespace
  \textbf{Other} & 659 & 0.75 \\
  \bottomrule
 \end{tabular}}
\end{table}

\section{\blue{PII Audit}}
\label{sec:pii-audit}

\blue{We ran openai/privacy-filter and eight credential-format patterns (\texttt{scripts/pii\_audit.py}) over the \texttt{prompt}, \texttt{content}, \texttt{schema}, and \texttt{response} fields of tag v1.0. Table~\ref{tab:pii-audit} counts distinct rows with at least one detection, per field and over any field. Person-name detections include any name on a public page (bylines, public figures, business owners). Date detections include publication dates.}

\begin{table}[h]
 \caption{\blue{Distinct rows with at least one PII detection (N=93,695).}}
 \label{tab:pii-audit}
 \small
 \centering
 \blue{\begin{tabular}{@{}lrrrrrr@{}}
  \toprule
  \textbf{PII class} & \textbf{prompt} & \textbf{content} & \textbf{schema} & \textbf{response} & \textbf{any field} & \textbf{\% of rows} \\
  \midrule
  Person name & 3,051 & 37,734 & 726 & 4,189 & 40,587 & 43.3 \\
  Private URL & 5,652 & 30,092 & 20 & 1,432 & 33,416 & 35.7 \\
  Secret or key & 25 & 8,710 & 10 & 56 & 8,754 & 9.3 \\
  Account number & 2,507 & 3,741 & 0 & 335 & 6,402 & 6.8 \\
  Phone & 677 & 4,216 & 7 & 942 & 5,279 & 5.6 \\
  Date & 755 & 3,595 & 82 & 1,467 & 5,247 & 5.6 \\
  Email & 52 & 4,707 & 8 & 912 & 5,140 & 5.5 \\
  Address & 128 & 2,781 & 8 & 1,388 & 3,933 & 4.2 \\
  \bottomrule
 \end{tabular}}
\end{table}

\blue{The secret class combines 8,732 rows flagged by the model with 39 rows matched by the credential patterns, and the two groups overlap. All 39 pattern-matched rows are in \texttt{content}: 16 rows with JWTs, 11 with AWS-format keys, 8 with Google-format keys, and 4 with OpenAI-format keys. The audit drops pattern hits that fall inside a model span, so 39 is an undercount. We rematched the patterns against the stored 80-character model snippets and found 13 more rows with AWS-format or Google-format keys. These rows are also in \texttt{content}, which gives at least 52 rows. We cannot rematch longer strings, such as JWTs, from snippets.}

\blue{We did not test whether any of these strings are valid credentials. The audit did not cover \texttt{source} or \texttt{llm\_model}. Dataset viewer statistics show identifiers in \texttt{llm\_model}. Three rows contain an AWS resource name with an account number, and 1 row contains an internal IP address.}

\blue{A reviewer flagged a fitness studio example. The audit finds 48 rows that reference this studio. The business published the studio's address, role-based email, booking phone, and trainer names on its own pages. We treat such spans as public business listings, retain them as benchmark ground truth, and remove them on request through the takedown process.}

\section{\blue{Domain Coverage}}
\label{sec:domain-diversity}

\blue{We derived each example's hostname from its source URL. We then collapsed hostnames to registered root domains (eTLD+1) with the Public Suffix List (\texttt{scripts/domain\_diversity.py}). The dataset spans 80,213 unique URLs, 10,295 hostnames, and 8,155 root domains. Of all examples, 99.9\% come from resolvable public web URLs.}

\blue{The top-10 domains account for 44.2\% of examples, and the top-100 domains account for 69.2\%. Of all domains, 56.3\% (4,588) contribute only one example. There are 61,858 unique schema-and-domain pairs, and 35.4\% of schemas appear on only one domain. Table~\ref{tab:top-domains} counts schemas by hash, so a template that embeds a book title counts once per title (Section~\ref{sec:processing}).}

\begin{table}[h]
 \caption{\blue{Top-10 root domains. Shares use the rows with a resolvable URL.}}
 \label{tab:top-domains}
 \small
 \centering
 \blue{\begin{tabular}{@{}lrrr@{}}
  \toprule
  \textbf{Domain} & \textbf{Examples} & \textbf{Share (\%)} & \textbf{Schema hashes} \\
  \midrule
  amazon.com & 10,155 & 10.9 & 8,381 \\
  google.com & 5,385 & 5.8 & 4,823 \\
  oup.com & 5,198 & 5.6 & 1,664 \\
  sagepub.com & 4,557 & 4.9 & 1,616 \\
  routledge.com & 3,595 & 3.8 & 1,693 \\
  vitalsource.com & 2,790 & 3.0 & 2,163 \\
  pearson.com & 2,607 & 2.8 & 931 \\
  wiley.com & 2,605 & 2.8 & 1,100 \\
  taylorfrancis.com & 2,454 & 2.6 & 1,116 \\
  cambridge.org & 1,976 & 2.1 & 768 \\
  \bottomrule
 \end{tabular}}
\end{table}

\section{\blue{Effect of Target Regeneration}}
\label{sec:regeneration}

\blue{We compared original responses with GPT-5-nano regenerations on the v1.0 snapshots (\texttt{scripts/regeneration\_comparison.py}). Both sets are 100\% parseable JSON. Schema compliance rises from 83.7\% to 100\%, because \texttt{modelling/regenerate.py} discards regenerations that fail validation. Regeneration largely preserves structure: top-level key Jaccard is 0.91, flattened-key Jaccard is 0.80 (median 1.0), and 73.4\% of pairs have identical flattened key sets. Regenerated targets are slightly shorter (median 511 vs.\ 565 characters, mean 6.7 vs.\ 7.3 leaf fields). Originals mark missing fields with the string ``NA'', while regenerations use null or empty values.}

\blue{At the value level, 74.4\% of shared leaves differ under exact comparison. The targets are identical in 7.0\% of pairs. The changed leaves fall into five groups:}
\begin{itemize}
 \item \blue{22.5\% differ only in the missing-value convention.}
 \item \blue{15.3\% fill a field that the original left missing.}
 \item \blue{13.8\% leave missing a field that the original filled.}
 \item \blue{0.4\% involve case, whitespace, or numeric formatting.}
 \item \blue{48.0\% are content rewrites.}
\end{itemize}

\blue{Table~\ref{tab:rewrite-taxonomy} splits the 32,458 content rewrites into classes (\texttt{scripts/rewrite\_taxonomy.py}). Deterministic rules classify 17.4\% of them as formatting, reordering, or strict containment. An LLM judge (\texttt{gpt-5.6-terra}) compares a seeded random sample of 1,475 of the remaining 26,814 leaves. The judged shares are therefore sample estimates, and we have not validated them against human labels. Over all changed shared leaves, this gives about 5\% meaning-preserving rewrites, 23\% partial changes, and 19\% substantive changes. Regeneration is thus mostly relabeling rather than rewording.}

\begin{table}[h]
 \caption{\blue{Split of the 32,458 content rewrites.}}
 \label{tab:rewrite-taxonomy}
 \small
 \centering
 \blue{\begin{tabular}{@{}lrl@{}}
  \toprule
  \textbf{Class} & \textbf{Share (\%)} & \textbf{Meaning} \\
  \midrule
  Meaning-preserving & 11.3 & Formatting, reordering, judged-equivalent paraphrases \\
  Partial overlap & 48.4 & Truncations completed, details added or dropped \\
  Substantive & 39.8 & Different entity, number, date, or fact \\
  Unclear & 0.5 & Judge abstentions, mostly URL or asset identifiers \\
  \bottomrule
 \end{tabular}}
\end{table}

\section{\blue{Frontier Structured-Output Baselines}}
\label{sec:frontier}

\blue{We ran \texttt{scripts/frontier\_baselines.py} on the fine-tuning test split after an 8192-token prompt filter, which leaves N=2,761. Table~\ref{tab:finetune-results} comes from a separate evaluation run with N=2,714. Our model's compliance is 0.919 in this run and 0.917 in the Table~\ref{tab:finetune-results} run. GPT-5.6 (\texttt{gpt-5.6-terra}) and Claude Sonnet 4.6 (\texttt{claude-sonnet-4.6}) used native \texttt{json\_schema} mode with \texttt{strict} set to false. Both models used the same prompt, 8192 maximum new tokens, provider-default temperature, and 16 concurrent calls. Our model used plain generation at temperature 0 with repetition penalty 1.1 and no constrained decoding.}

\blue{When a provider rejected a native-mode request, the script retried the call in plain \texttt{json\_object} mode (the fallback), so every completed call has a scored prediction. Most rejections concern the schema, but provider limits and content-policy errors also trigger the fallback (see below). We compute compliance and key F1 in Table~\ref{tab:frontier} over all completed calls. GPT-5.6 used the fallback on 3 calls: 2 invalid schemas and 1 content-policy error. In the 100-row \texttt{human\_eval} run, Claude used the fallback on 55 calls (45 rate-limited, 10 rejected).}

\begin{table}[h]
 \caption{\blue{Compliance and key F1 over all completed calls. Claude completed 2,760 of 2,761 calls. Groups 0--50 to 1000+ are composite complexity-score buckets.}}
 \label{tab:frontier}
 \small
 \centering
 \blue{\begin{tabular}{@{}lrcccccc@{}}
  \toprule
  & & \multicolumn{2}{c}{\textbf{GPT-5.6}} & \multicolumn{2}{c}{\textbf{Claude Sonnet 4.6}} & \multicolumn{2}{c}{\textbf{Ours (1.7B)}} \\
  \textbf{Group} & \textbf{N} & Compl. & Key F1 & Compl. & Key F1 & Compl. & Key F1 \\
  \midrule
  Overall & 2,761 & 0.988 & 0.929 & 0.983 & 0.915 & 0.919 & 0.887 \\
  0--50 & 493 & 0.996 & 0.977 & 0.990 & 0.970 & 0.955 & 0.937 \\
  50--100 & 905 & 0.992 & 0.938 & 0.981 & 0.927 & 0.950 & 0.907 \\
  100--200 & 892 & 0.991 & 0.919 & 0.990 & 0.904 & 0.910 & 0.872 \\
  200--500 & 347 & 0.977 & 0.905 & 0.980 & 0.865 & 0.865 & 0.841 \\
  500--1000 & 92 & 0.957 & 0.861 & 0.912 & 0.863 & 0.783 & 0.826 \\
  1000+ & 32 & 0.844 & 0.650 & 1.000 & 0.726 & 0.719 & 0.625 \\
  Depth $\geq$ 7 & 408 & 0.980 & 0.841 & 0.968 & 0.818 & 0.853 & 0.795 \\
  Keys $\geq$ 200 & 72 & 0.903 & 0.732 & 0.931 & 0.777 & 0.722 & 0.708 \\
  \bottomrule
 \end{tabular}}
\end{table}

\blue{Of the 1,704 fallbacks in the Claude run, 1,459 came from a provider limit of 20 grammar compilations per minute, which our 16-way concurrency exceeded. These rate-limited calls say nothing about the schema. An earlier report of this run gave a 62\% schema rejection rate. There are 245 genuine rejections: at least 8.9\% of all calls and 18.8\% of the 1,301 calls with a verdict. The genuine-rejection rate rises from 4.5\% of calls with a verdict in the simplest bucket to 91\% in the most complex. The genuine rejections have four causes: unsupported \texttt{additionalProperties} forms (89), schema size or compilation limits (132), other unsupported keywords (14), and degenerate user schemas (10).}

\blue{Table~\ref{tab:rejection} reports two end-to-end compliance scores. The \emph{Strict} score counts every fallback as a failure, so the rate limit dominates it. The \emph{Genuine} score counts only genuine schema rejections as failures and keeps the scored fallback for rate-limited calls. It is therefore an upper bound on native-mode success, not a measurement of it. GPT-5.6's compliance of 0.988 is close to its end-to-end score because only 3 of its calls used the fallback.}

\begin{table}[h]
 \caption{\blue{Claude Sonnet 4.6 rejection accounting. Accepted compliance uses accepted calls only. Groups are the same as in Table~\ref{tab:frontier}.}}
 \label{tab:rejection}
 \small
 \centering
 \blue{\begin{tabular}{@{}lrrrrccc@{}}
  \toprule
  \textbf{Group} & \textbf{Done} & \textbf{Accepted} & \textbf{Rate-limited} & \textbf{Rejected} & \textbf{Accepted compl.} & \textbf{Strict} & \textbf{Genuine} \\
  \midrule
  Overall & 2,760 & 1,056 & 1,459 & 245 & 0.999 & 0.382 & 0.899 \\
  0--50 & 493 & 211 & 272 & 10 & 1.000 & 0.428 & 0.970 \\
  50--100 & 905 & 371 & 518 & 16 & 1.000 & 0.410 & 0.966 \\
  100--200 & 892 & 368 & 492 & 32 & 0.997 & 0.411 & 0.955 \\
  200--500 & 347 & 100 & 161 & 86 & 1.000 & 0.288 & 0.735 \\
  500--1000 & 91 & 3 & 16 & 72 & 1.000 & 0.033 & 0.209 \\
  1000+ & 32 & 3 & 0 & 29 & 1.000 & 0.094 & 0.094 \\
  Depth $\geq$ 7 & 407 & 96 & 174 & 137 & 1.000 & 0.236 & 0.649 \\
  Keys $\geq$ 200 & 72 & 3 & 2 & 67 & 1.000 & 0.042 & 0.069 \\
  \bottomrule
 \end{tabular}}
\end{table}

\section{\blue{Human-Verified Evaluation Protocol}}
\label{sec:human-protocol}

\blue{The \texttt{human\_eval} split (fine-tuning tag v1.1) holds 100 examples sampled with a fixed seed from the test split. We removed test rows that share a chunk, a page, or a schema-and-response pair with training. We also removed rows whose provenance we could not align. This leaves 527 rows, and the 8192-token filter leaves 525. We stratified the sample by schema-complexity bucket and by full or truncated page (87 full, 13 truncated). \texttt{scripts/human\_benchmark.py} reproduces the sample.}

\blue{Each example started from the GPT-5-nano draft. Three of the authors annotated every example. They verified or corrected every leaf value against the source page content. Annotators saw the schema and the exact content that the model saw. Annotators followed these rules:}
\begin{itemize}
 \item \blue{Correct wrong or paraphrased values by copying from the content.}
 \item \blue{Fill a missing value when the content contains it.}
 \item \blue{Replace values that the content does not support with null.}
 \item \blue{Use null when the content lacks a value.}
 \item \blue{Use no outside knowledge.}
 \item \blue{Keep each record schema-compliant.}
 \item \blue{Note ambiguous cases.}
\end{itemize}

\end{document}